\documentclass{aa}  
\usepackage{graphicx}
\usepackage{txfonts}
\newcommand{\eg}{{\it e.g.}, }
\newcommand{\ie}{{\it i.e.}, }
\newcommand{\ms}{m\,s$^{\rm -1}$}
\newcommand{\kms}{km\,s$^{\rm -1}$}
\newcommand{\Mjup}{M$_{\rm Jup}$}
\newcommand{\Msun}{M$_{\sun}$}
\newcommand{\vsini}{$v\sin{i}$}
\newcommand{\elodie}{E{\small LODIE}}
\newcommand{\sophie}{S{\small OPHIE}}
\newcommand{\harps}{H{\small ARPS}}
\newcommand{\thetacyg}{$\theta$\,Cygni}

\begin{document}
   \title{Extrasolar planets and brown dwarfs around A--F type stars
     \thanks{Based on observations made with the \elodie~and \sophie~spectrographs at the Observatoire de Haute-Provence (CNRS, France) and with the PUEO adaptive optics system at the Canada-France-Hawaii Telescope (CFHT) which is operated by the National Research Council of Canada, the Institut National des Sciences de l'Univers of the Centre National de la Recherche Scientifique of France, and the University of Hawaii.}
     \thanks{Tables of radial velocities are only available in electronic form at the CDS via anonymous ftp to cdsarc.u-strasbg.fr (130.79.128.5) or via http://cdsweb.u-strabg.fr/cgi-bin/qcat?J/A+A/}}

   \subtitle{VII. \object{\thetacyg}~radial velocity variations: planets or stellar phenomenon?}

   \author{
     M. Desort \inst{1} \and
     A.-M. Lagrange \inst{1} \and
     F. Galland \inst{1} \and
     S. Udry \inst{2} \and
     G. Montagnier \inst{2,1} \and
     H. Beust \inst{1} \and
     I. Boisse \inst{3} \and
     X. Bonfils \inst{1,4} \and
     F. Bouchy \inst{3} \and
     X. Delfosse \inst{1} \and
     A. Eggenberger \inst{1} \and
     D. Ehrenreich  \inst{1} \and
     T. Forveille \inst{1} \and
     G. H\'ebrard \inst{3} \and
     B. Loeillet \inst{3,5} \and
     C. Lovis \inst{2} \and
     M. Mayor \inst{2} \and
     N. Meunier \inst{1} \and
     C. Moutou \inst{5} \and
     F. Pepe \inst{2} \and
     C. Perrier \inst{1} \and
     F. Pont \inst{6} \and 
     D. Queloz \inst{2} \and
     N. C. Santos \inst{2,4} \and
     D. S\'egransan \inst{2} \and
     A. Vidal-Madjar \inst{3}
   }

   \institute{
     Laboratoire d'Astrophysique de Grenoble, UMR5571 CNRS, Universit\'e Joseph Fourier, BP 53, 38041 Grenoble Cedex 9, France\\
     \email{morgan.desort@obs.ujf-grenoble.fr}
     \and
     Observatoire de Gen\`eve, Universit\'e de Gen\`eve, 51 Chemin des Maillettes, 1290 Sauverny, Switzerland
     \and
     Institut d'Astrophysique de Paris, UMR7095 CNRS, Universit\'e Pierre \& Marie Curie, 98bis boulevard Arago, 75014 Paris, France
     \and
     Centro de Astronomia e Astrof{\'\i}sica da Universidade de Lisboa, Observat\'orio Astron\'omico de Lisboa, Tapada da Ajuda, 1349-018 Lisboa, Portugal
     \and
     Laboratoire d'Astrophysique de Marseille, UMR6110 CNRS, Universit\'e de Provence, BP 8, 13376 Marseille Cedex 12, France
     \and
     Physikalisches Institut, University of Bern, Sidlerstrasse 5, 3012 Bern, Switzerland 
   }

   \date{Received date / Accepted date}

   
   \abstract
   {}
   {In the frame of the search for extrasolar planets and brown dwarfs around
   early-type main-sequence stars, we present the results obtained on the early
   F-type star \thetacyg.
   }
   {\elodie~and \sophie~at Observatoire de Haute-Provence (OHP) were used to
   obtain 91 and 162 spectra, respectively. Our dedicated radial-velocity
   measurement method was used to monitor the star's radial velocities over five
   years. We also use complementary, high angular resolution and high-contrast
   images taken with PUEO at CFHT.
   }
   {We show that \thetacyg~radial velocities are quasi-periodically variable,
   with a $\simeq$ 150-day period. These variations are not due to the
   $\simeq$\,0.35-\Msun~stellar companion that we detected in imaging at more
   than 46\,AU from the star.

   The absence of correlation between the bisector velocity span variations and
   the radial velocity variations for this 7\,\kms~\vsini~star, as well as other
   criteria indicate that the observed radial velocity variations are not due to
   stellar spots. The observed amplitude of the bisector velocity span
   variations also seems to rule out stellar pulsations. However, we observe a
   peak in the bisector velocity span periodogram at the same period as the one
   found in the radial velocity periodogram, which indicates a probable link
   between these radial velocity variations and the low amplitude lineshape
   variations which are of stellar origin. Long-period variations are not
   expected from this type of star to our knowledge. If a stellar origin (hence
   of new type) was to be confirmed for these long-period radial velocity
   variations, this would have several consequences on the search for planets
   around main-sequence stars, both in terms of observational strategy and data
   analysis.

   An alternative explanation for these variable radial velocities is the
   presence of at least one planet of a few Jupiter masses orbiting at less than
   1\,AU; however this planet alone does not explain all observed features, and
   the \thetacyg~system is obviously more complex than a planetary system with 1
   or 2 planets.
   }
   {The available data do not allow to distinguish between these two possible
   origins. A vigourous follow-up in spectroscopy and photometry is needed to
   get a comprehensive view of the star intrinsic variability and/or its
   surrounding planetary system.
   }

   \keywords{techniques: radial velocities - stars: early-type - stars: planetary systems - stars: individual: \thetacyg}

   \maketitle

%

\section{Introduction}

Radial-velocity (RV) surveys have lead to the detection of more than 300 planets
during the past decade\footnote{A comprehensive list of known exoplanets is
available at http://exoplanet.eu}. These surveys mainly focus on solar and
later-type main-sequence (hereafter MS) stars ($\ga$ F7) which exhibit numerous
lines with low rotational broadening, making them ideal targets for classical
velocimetry. However, it is crucial to understand how planetary systems form
over a wide variety of parent stars, and to know in particular if there is a
correlation between the planet masses and the parent star masses as predicted
for instance by \cite{kennedy08} (2008), and to constrain formation models such
as those from \cite{ida05} (2005) (but also \cite{boss06} 2006 and
\cite{laughlin04} 2004 for M dwarfs), and/or a correlation between the planet
occurence and the parent star masses. Concerning massive stars, surveys of
subgiant/giant stars have started to provide first information on planets at
orbital distances typically greater than 0.7\,AU (\eg \cite{johnson06} 2006;
\cite{johnson07} 2007; \cite{hatzes05} 2005; \cite{niedzielski07} 2007;
\cite{lovis07} 2007; \cite{sato08} 2008). Closer separations have to be
investigated by observing massive main-sequence stars. In this frame, we
developed a tool dedicated to the search for planets around early (A--F) type
stars. The method allowing the measurement of the RV of rapid rotators is
described by \cite{galland05a} (2005a; hereafter Paper\,I).

We started in 2005 two surveys dedicated to the search for extrasolar planets
and brown dwarfs around a volume-limited sample of A--F main-sequence stars {\it
i)} with the \elodie~fiber-fed echelle spectrograph (\cite{baranne96} 1996)
mounted on the 1.93-m telescope at the Observatoire de Haute-Provence (OHP,
France) in the northern hemisphere, and {\it ii)} with the \harps~spectrograph
(\cite{pepe02} 2002) installed on the 3.6-m ESO telescope at La Silla
Observatory (Chile) in the southern hemisphere. In 2006, the
\elodie~spectrograph was replaced by \sophie~(\cite{bouchy06} 2006). We detected
with \elodie~a planet around an F6V star (\cite{galland05b} 2005b, Paper\,II)
and a brown dwarf around an A9V star (\cite{galland06b} 2006b, Paper\,IV), and
with \harps~a two-planet system around an F6IV--V star (\cite{desort08} 2008,
Paper\,V). We also derived the first statistics of planet existence around A--F
stars thanks to our \harps~survey (\cite{lagrange09} 2009, Paper\,VI).

We present and analyse in this paper the RV variations of \thetacyg.
Section~\ref{stellar_param} provides the stellar properties, and the various
data obtained on this object. In Sect.\,\ref{keplerian_sol}, we discuss the
origin of the observed RV variations.

\section{Stellar characteristics and measurements}
\label{stellar_param}

  \subsection{Stellar properties}

  \thetacyg~(\object{HD\,185395}, \object{HIP\,96441}, \object{HR\,7469}) is a
  $M_1=1.38\pm0.05$\,\Msun~star, with an age estimated to
  $1.5_{-0.7}^{+0.6}$\,Gyr (\cite{nordstrom04} 2004), and located at
  $18.33\pm0.05$\,pc from the Sun (\cite{hipparcos97} 1997, \cite{vanleeuwen07}
  2007). We took its rotational velocity \vsini, effective temperature $T_{\rm
  eff}$, and surface gravity $\log{g}$ from \cite{erspamer03} (2003) and
  \cite{gray03} (2003) (values in Table~\ref{Table_hd185395_stelpar}). We assume
  a spectral type F4V, commonly attributed to this star as, \eg in the Bright
  Star Catalogue (\cite{hoffleit91} 1991) or in the H{\small IPPARCOS} catalogue
  (\cite{hipparcos97} 1997).



  \begin{table}[t!]
    \caption{\thetacyg~stellar properties. Photometric and astrometric data are extracted from the H{\small IPPARCOS} catalogue (\cite{hipparcos97} 1997, \cite{vanleeuwen07} 2007); spectroscopic data are from \cite{nordstrom04} (2004) and \cite{erspamer03} (2003).}
    \label{Table_hd185395_stelpar}
    \begin{center}
      \begin{tabular}{l l c}
        \hline
	\hline
        Parameter       &                      & \thetacyg  \\
	\hline
        Spectral Type   &                      & F4V \\
        \vsini          & [\kms]               & 7  \\
        $V$             &                      & 4.49 \\
        $B-V$           &                      & $0.395 \pm 0.015$ \\
        $\pi$           & [mas]                & $54.54 \pm 0.15$  \\
        Distance        & [pc]                 & $18.33\pm 0.05$ \\
        $M_V$           &                      & 3.14 \\
	$[$Fe/H$]$      &                      & $-0.08$ \\
        $T_{\rm eff}$   & [K]                  & 6745 \\
        $\log{g}$       &                      & 4.2  \\
        $M_1$           & [\Msun]              & $1.38 \pm 0.05$ \\
	Age             & [Gyr]                & $1.5_{-0.7}^{+0.6}$  \\
	ppm ($\alpha$)$^\dagger$ & [mas\,yr$^{\rm -1}$] & $-8.15$\\
	ppm ($\delta$)$^\dagger$ & [mas\,yr$^{\rm -1}$] & $-262.99$\\
        \hline
      \end{tabular}
    \end{center}
    \begin{list}{}{}
    \item[$^\dagger$] the proper motion is affected from the orbital motion that we discuss in Section~\ref{sec:ao}.
    \end{list}
  \end{table}

  \subsection{Spectroscopic data}
  \subsubsection{Description of the data}
 
  Between 2003 and 2006, we recorded 91 high $\mathrm{S/N}$ spectra of
  \thetacyg~with \elodie~and, between November 2006 and December 2008, we
  recorded 162 spectra with \sophie.


  The wavelength range is 3850--6800\,\AA~for \elodie~and 3872--6943\,\AA~for
  \sophie. Typical exposure times were 15 and 3\,min respectively for
  \elodie~and \sophie, leading to a signal-to-noise ratio $\mathrm{S/N}$
  $\sim$\,200. The exposures were performed with simultaneous-thorium spectra to
  follow and correct for the possible drift of the instrument due to local
  temperature/pressure variations (whose impact shows a standard deviation of
  2.5\,\ms). With \sophie~we used the high-resolution ($R\approx75\,000$) mode.

  \subsubsection{Radial velocity variations}
 
\begin{figure*}[t!]
    \centering
    \includegraphics[width=0.9\hsize]{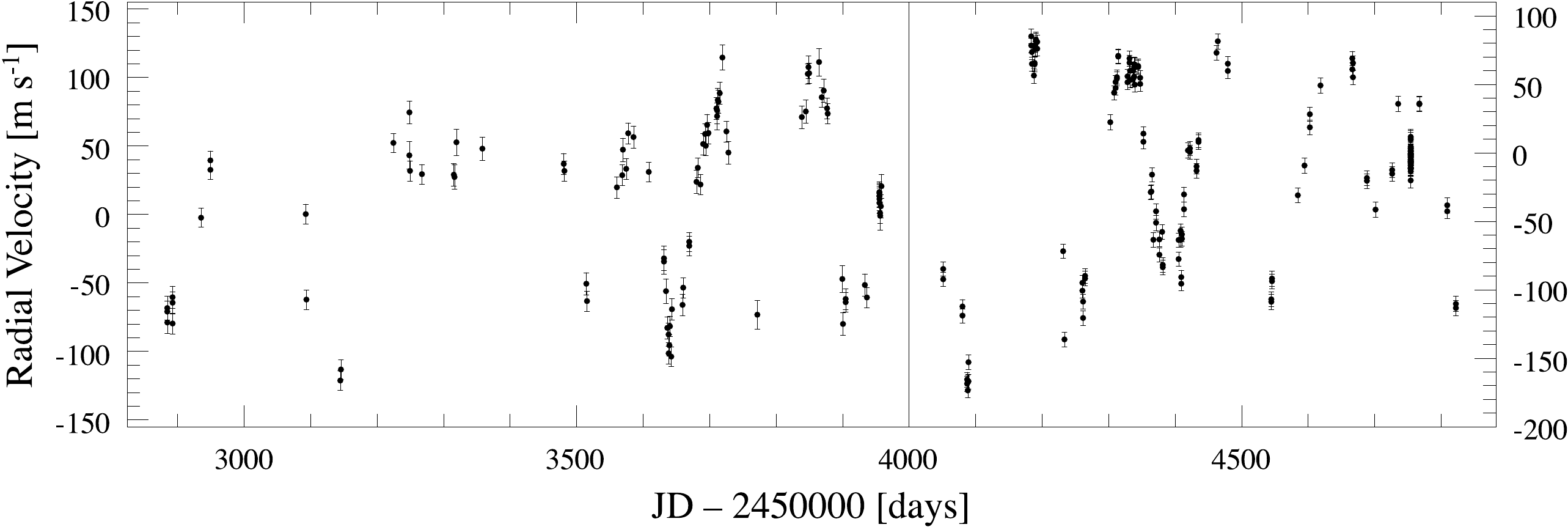}
    \caption{Radial velocities of \thetacyg~obtained with \elodie~({\it left})
    and \sophie~({\it right}).}
    \label{rv}
  \end{figure*}

  The radial velocities (Fig.\,\ref{rv}) are measured using a dedicated tool
  (S{\small AFIR}) described in Paper\,I and based on the Fourier interspectrum
  method developed in \cite{chelli00} (2000). The uncertainty associated with
  \elodie~data is 9\,\ms~on average, consistent with the value obtained from our
  simulations (see Paper\,I). In the case of \sophie~data, the uncertainty is
  5\,\ms~on average (taking the photon noise and instrument stability into
  account).

  As \thetacyg~has a relatively low projected rotational-velocity
  (\vsini\,=\,7\,\kms), we could also measure the RV using a gaussian
  adjustement to the cross-correlation function (CCF). The results obtained by
  the two different methods are found to be consistent.

  \elodie~and \sophie~RV data show (Fig.\,\ref{rv}) a quasi-periodic signal with
  peak-to-peak amplitude of about 220\,\ms, much larger than the
  uncertainties. A drift in the RV curve seems moreover to be present over the
  whole data set, which could be attributed to a stellar companion
  (Section~\ref{sec:ao}). We finally note that the amplitude of the RV
  variations could in addition be slightly variable.

\begin{figure}[t!]
  \centering
  \includegraphics[width=1\hsize]{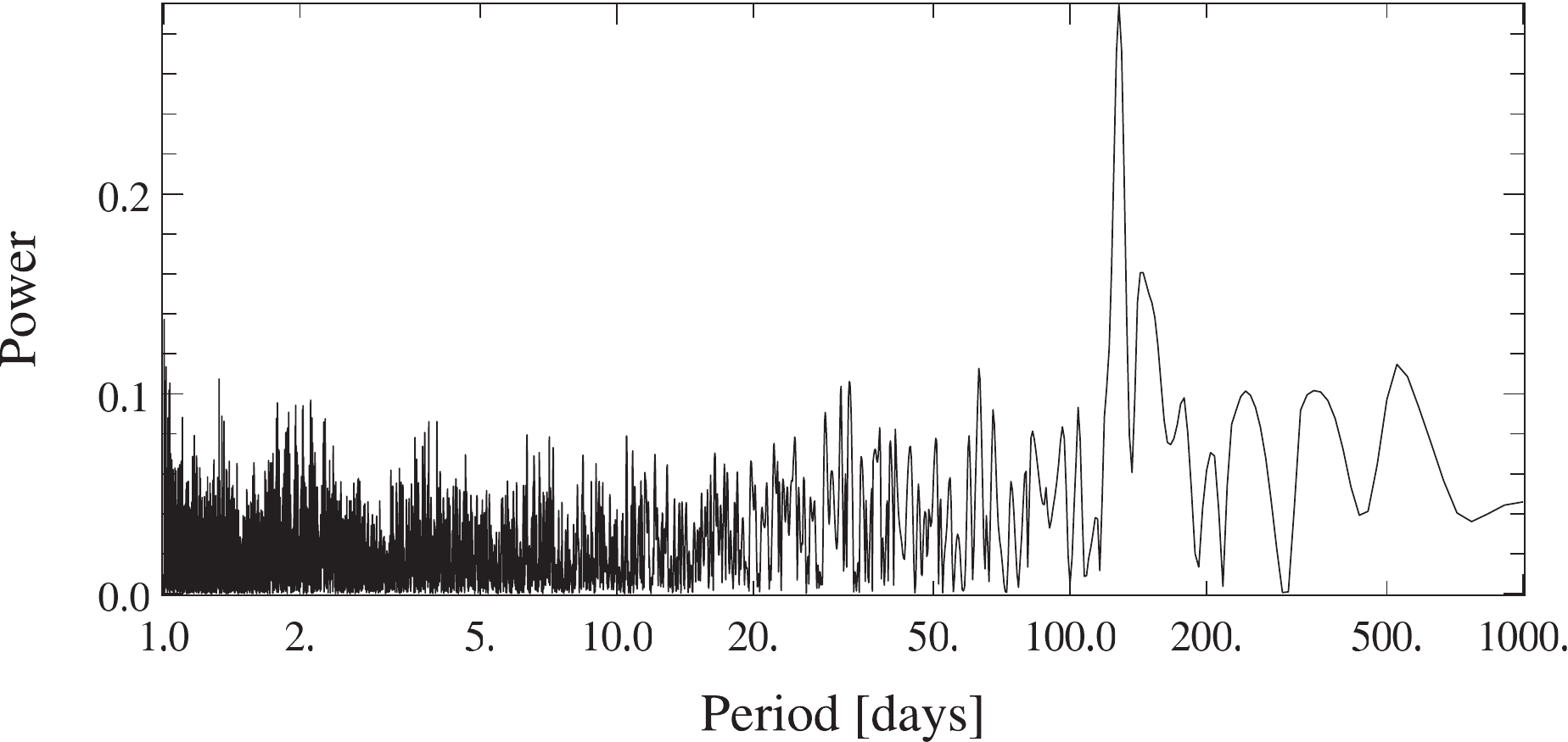}
  \includegraphics[width=1\hsize]{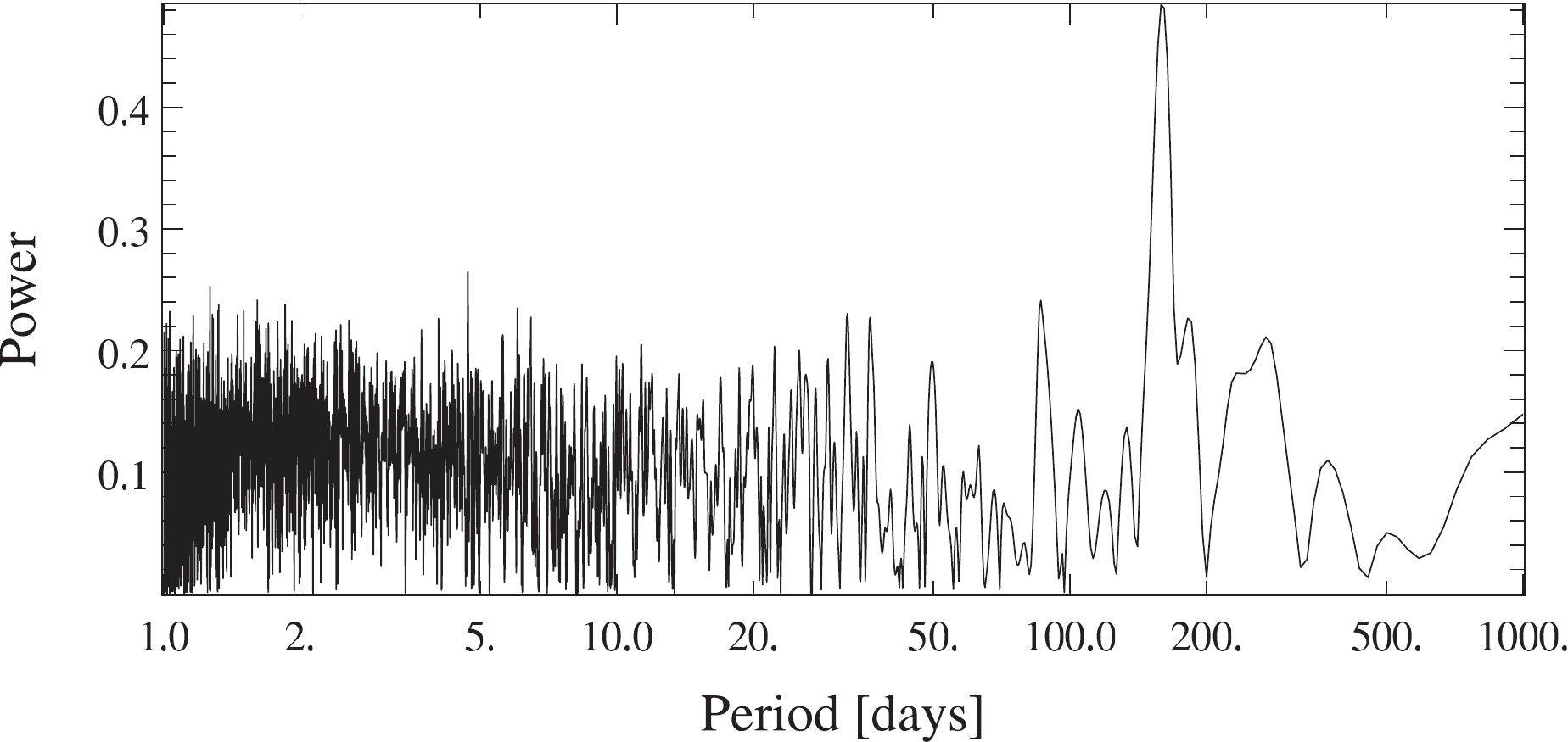}
  \caption{\elodie~({\it top}) and \sophie~({\it bottom}) CLEAN periodograms of
  the radial velocities. On each data set one peak is observed, either at $128
  \pm 5$ days or at $158 \pm 10$ days respectively.}
  \label{periodograms}
  \end{figure}

  We used the CLEAN algorithm (\cite{hogbom74} 1974), applied to Lomb-Scargle
  periodograms to derive the periodograms of the radial velocities measured with
  both instruments (Fig.\,\ref{periodograms}); this algorithm removes the
  aliases associated with temporal sampling of the data: it deconvolves the
  window function iteratively from the initial ``dirty'' periodogram to produce
  the resulting cleaned periodogram.

  In the case of \elodie~data, the periodogram shows one peak at a period of
  $128 \pm 5$ days (the uncertainty is evaluated with the full width at half
  maximum of the highest peak). In the case of \sophie~data, the peak
  corresponds to a period of $158 \pm 10$ days.

   \subsubsection{Line profile variations}

\begin{figure}[t!]
     \centering
     \includegraphics[width=0.7\hsize]{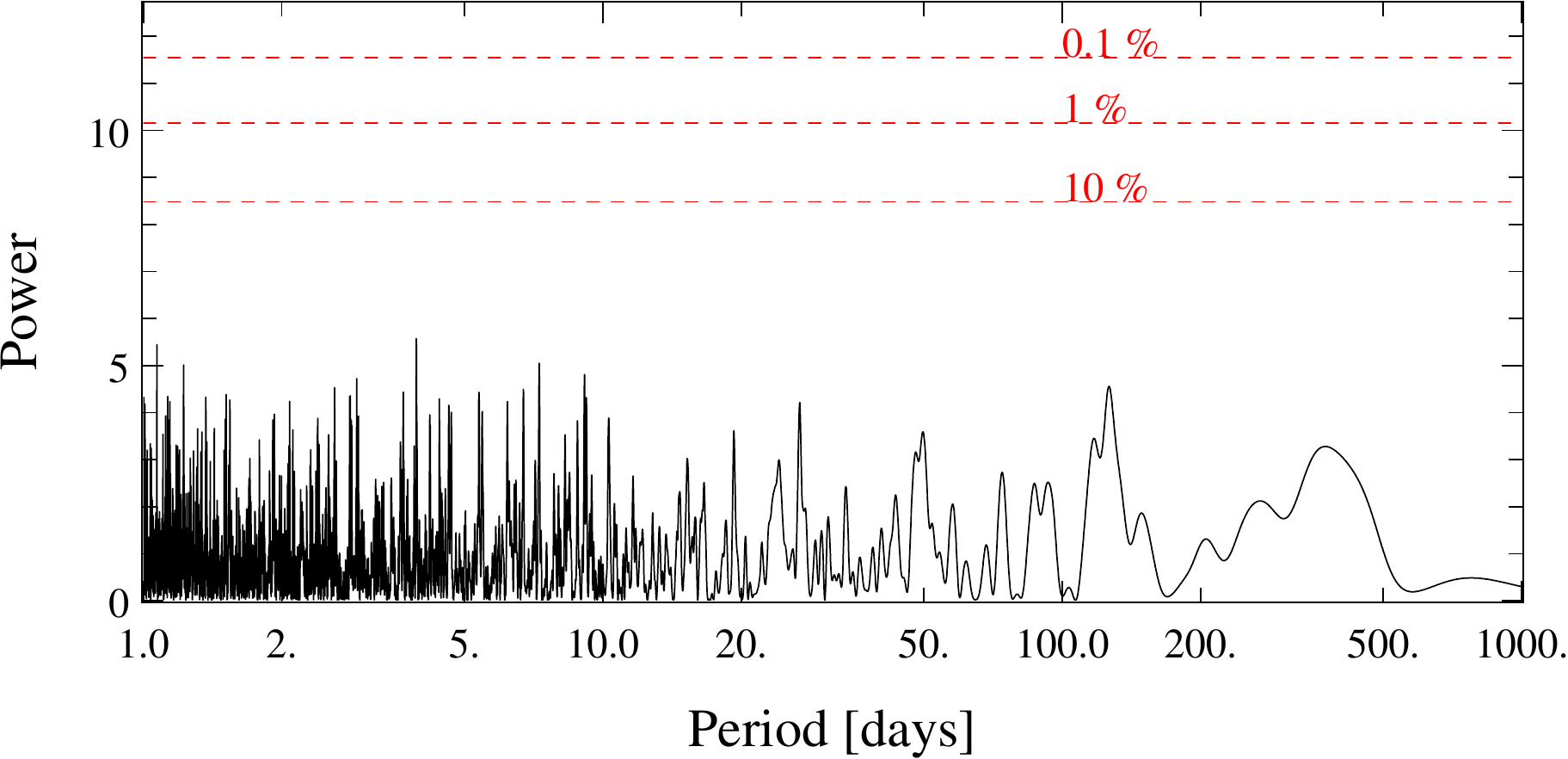}\vspace{0.2cm}
     \includegraphics[width=0.7\hsize]{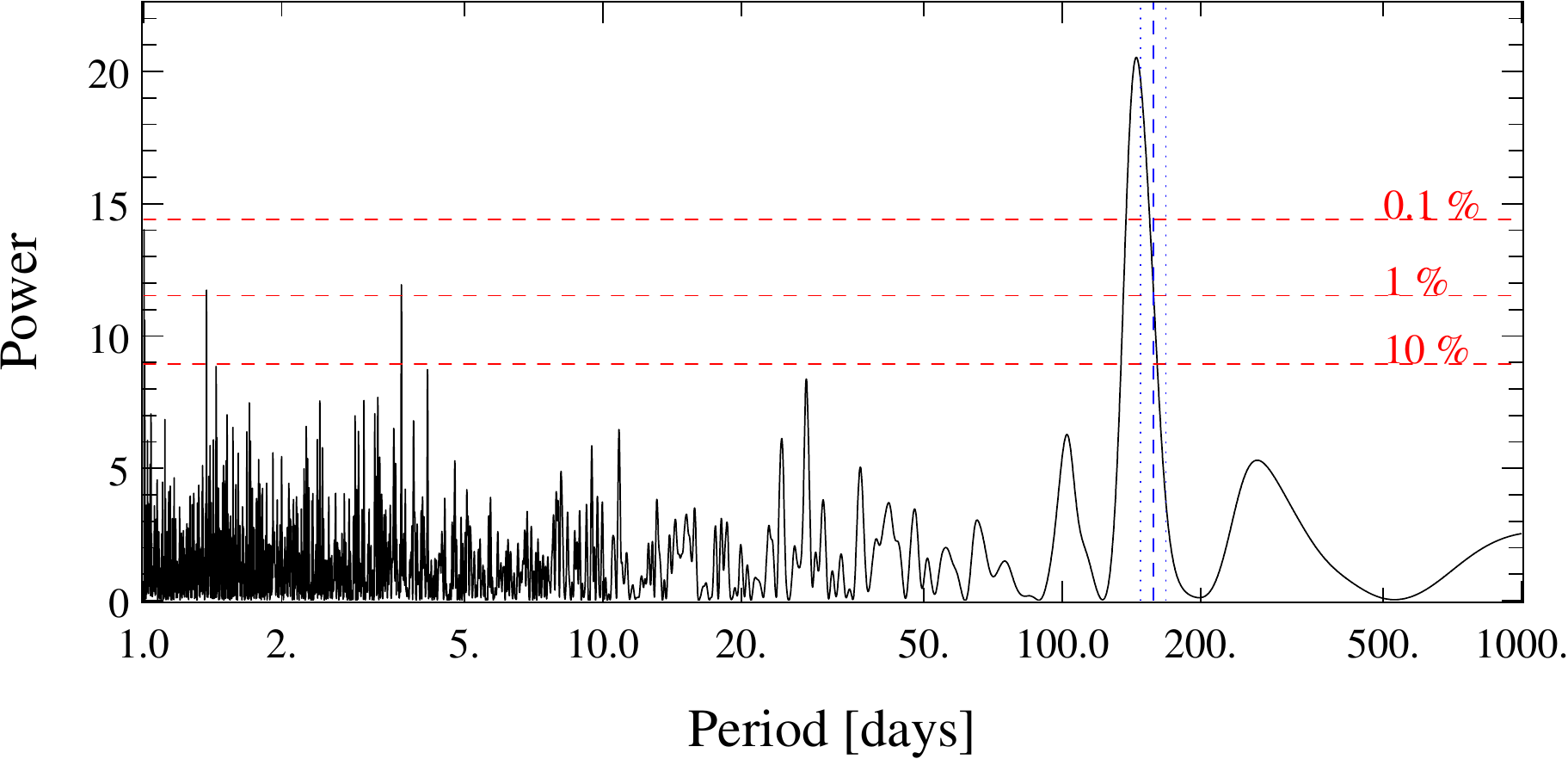}\vspace{0.2cm}
     \includegraphics[width=0.7\hsize]{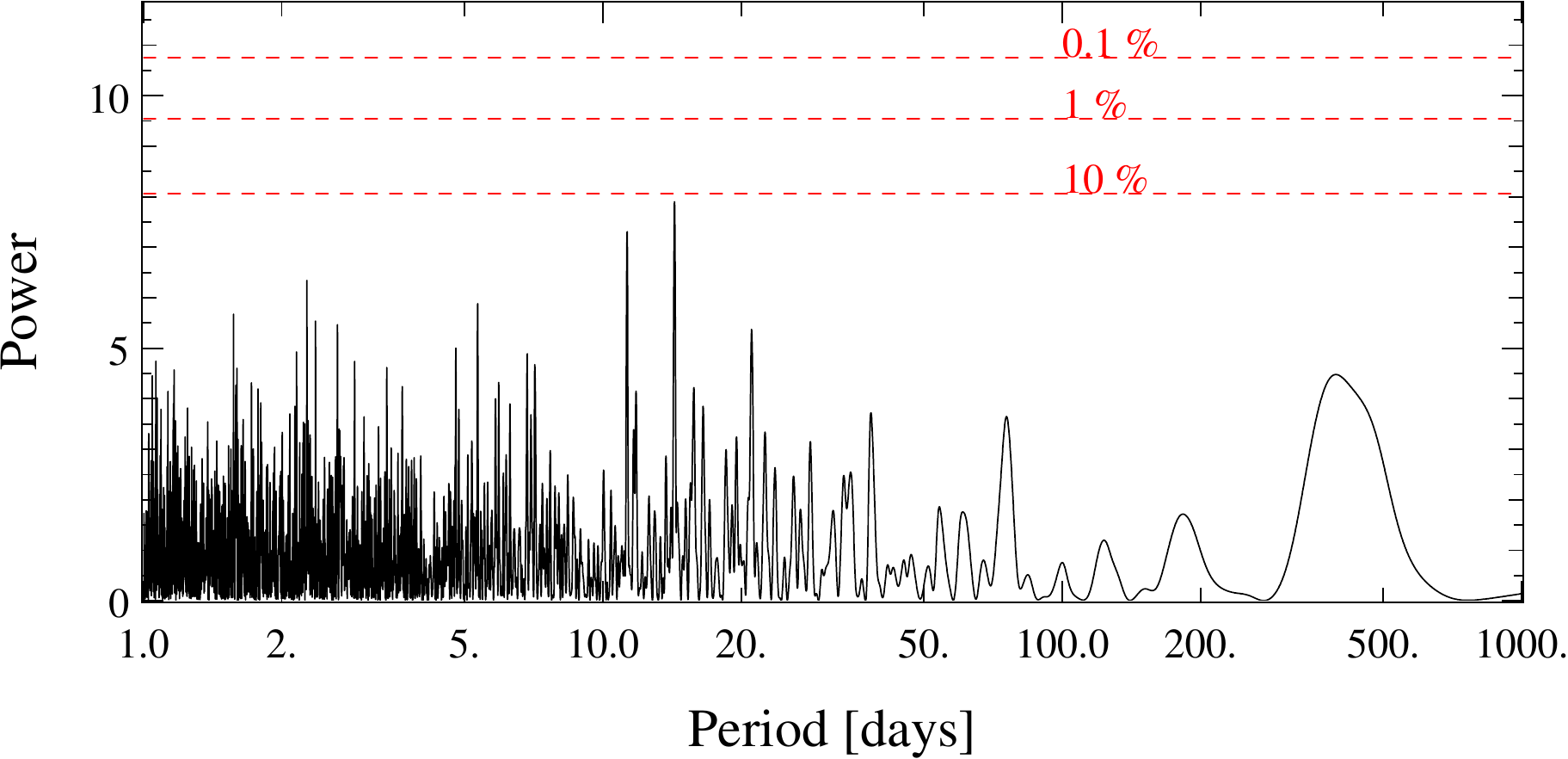}\vspace{0.2cm}
     \includegraphics[width=0.7\hsize]{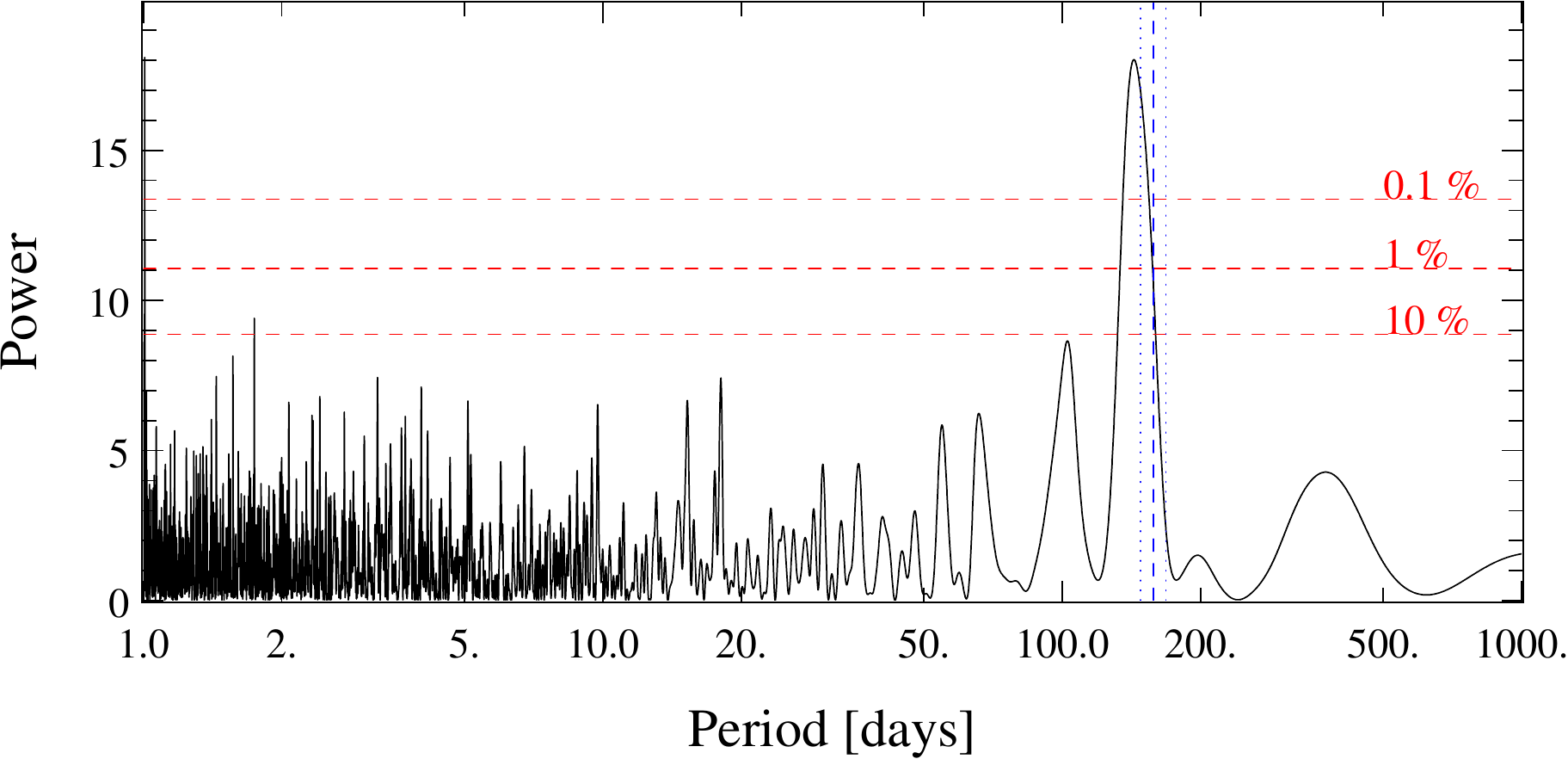}
     \caption{\elodie~and \sophie~Lomb-Scargle periodograms of the bisector
       velocity spans ({\it top 2 panels}) and curvatures ({\it bottom 2
       panels}) with false alarm probabilities. {\bf On \sophie periodograms the
       period of 158$\pm$10 days is represented with vertical dashed lines
       (dotted for the $\pm 10$ range).}}
     \label{span_curv_periodograms}
   \end{figure}

   \sophie~Lomb-Scargle periodograms of the BVSs and curvatures (defined as in
   \cite{hatzes96} 1996) are presented in Fig.\,\ref{span_curv_periodograms},
   together with false alarm probabilities (FAP, \cite{kurster97} 1997). In the
   case of the \sophie~data, a peak is seen at approximately 140~days, \ie not
   very different from the one measured in the \sophie~RV variations. No peak is
   detected on the yet noisier \elodie~data.

\begin{figure}[t!]
     \centering
     \includegraphics[width=0.8\hsize]{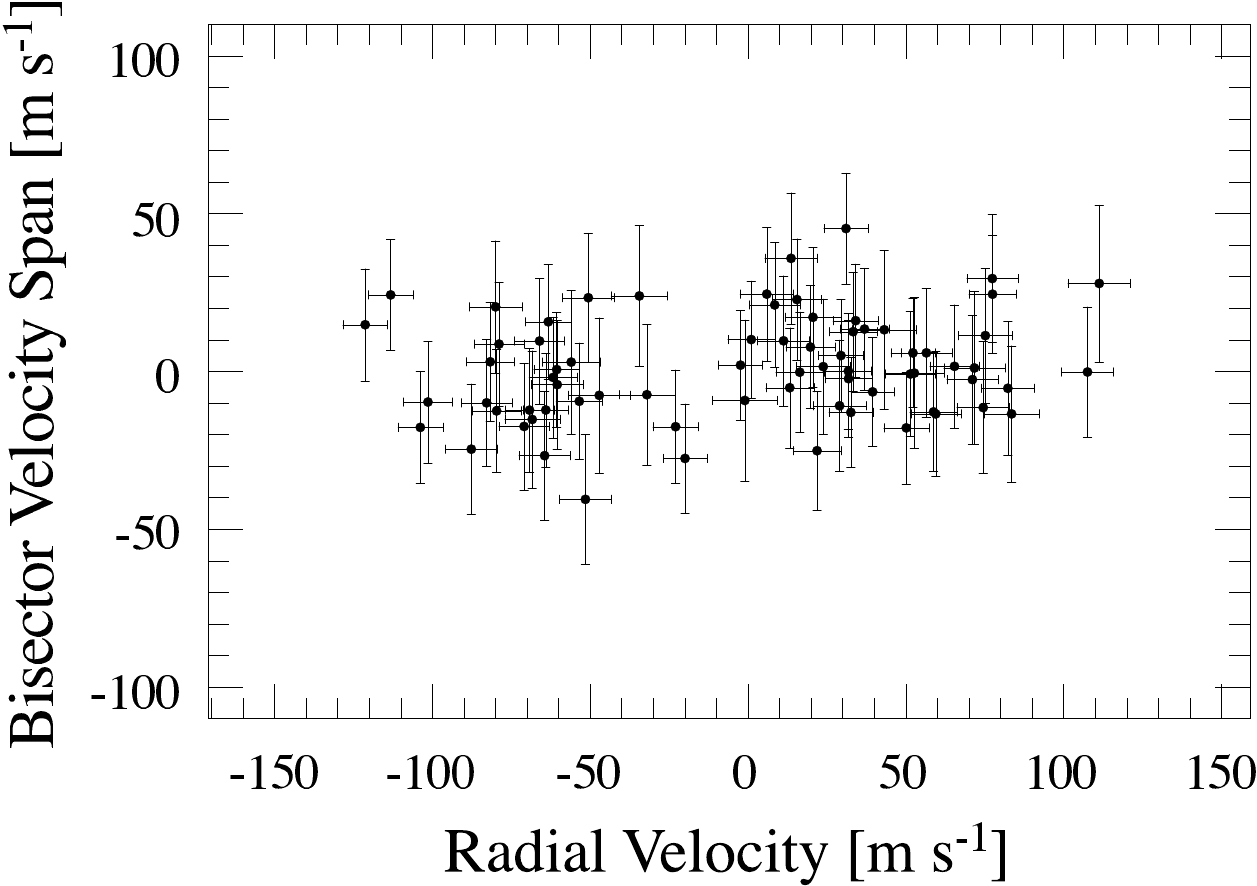}
     \includegraphics[width=0.8\hsize]{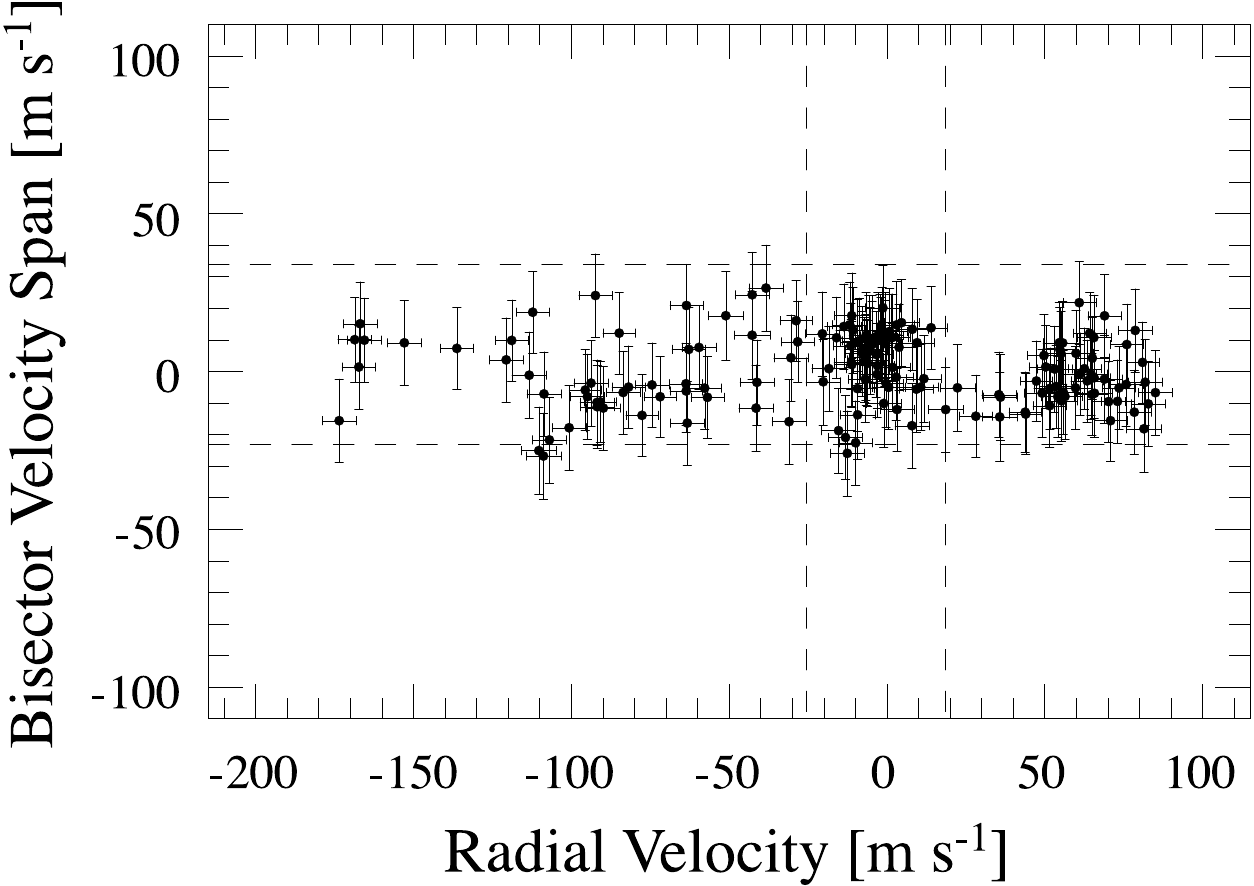}
     \caption{Bisector velocity span versus RVs for \elodie~({\it top}) and
       \sophie~({\it bottom}) data, showing that there is no correlation between
       line profiles and radial velocities. The dashed lines on \sophie~data
       show the minimum effect of short-term variations, see
       Section~\ref{short_term_var}.}
     \label{bis_span}
   \end{figure}

   Figure~\ref{bis_span} presents the bisector velocity spans (BVSs) as a
   function of RVs for \elodie~and \sophie~spectra. The amplitude of the BVS
   variations is quite small: 50\,\ms, much smaller than the amplitude of the RV
   variations. No clear correlation is seem between the BVS and RV
   variations. It shows that the spectra are mainly shifted in radial velocity
   without significant changes in the lines shape.

 \subsubsection{Stellar jitter}
 \label{short_term_var}

\begin{figure}[t!]
    \centering
    \includegraphics[width=0.8\hsize]{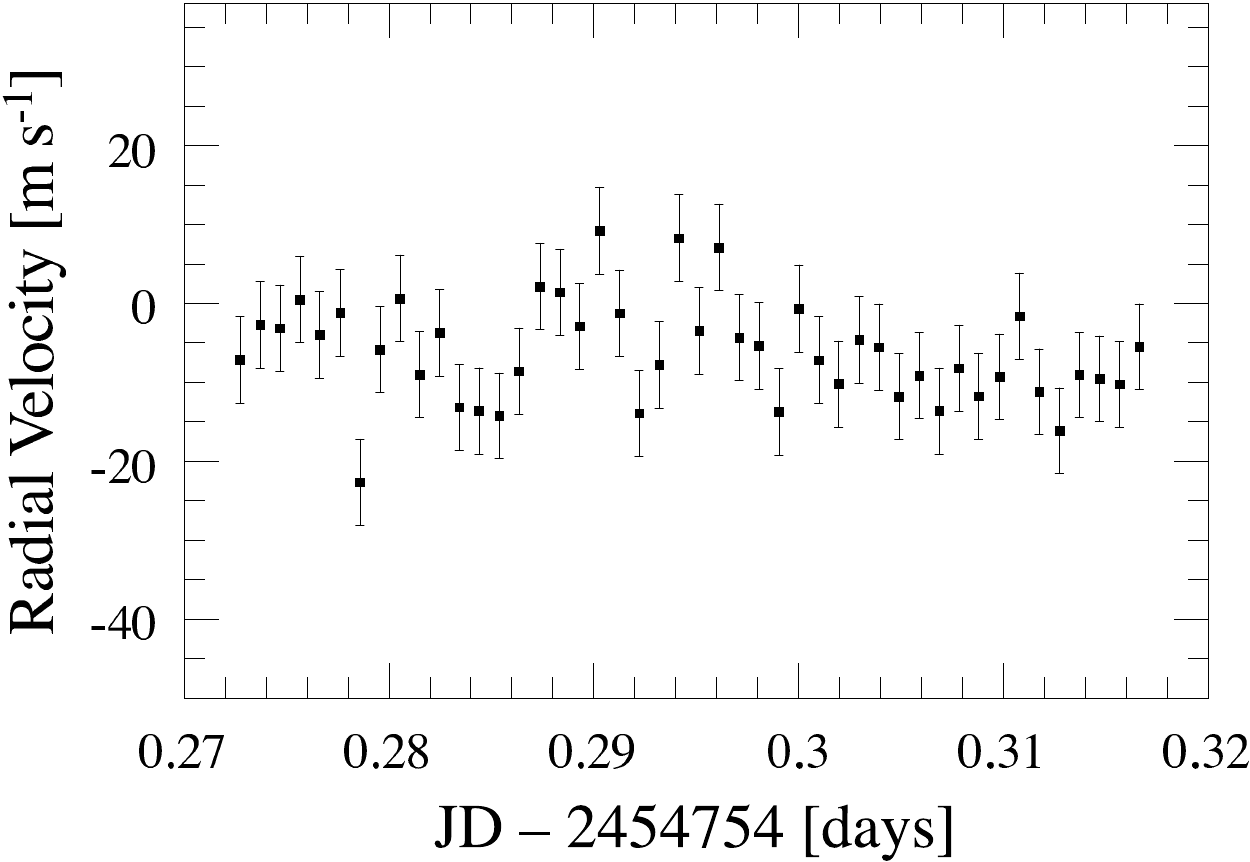}
    \includegraphics[width=0.8\hsize]{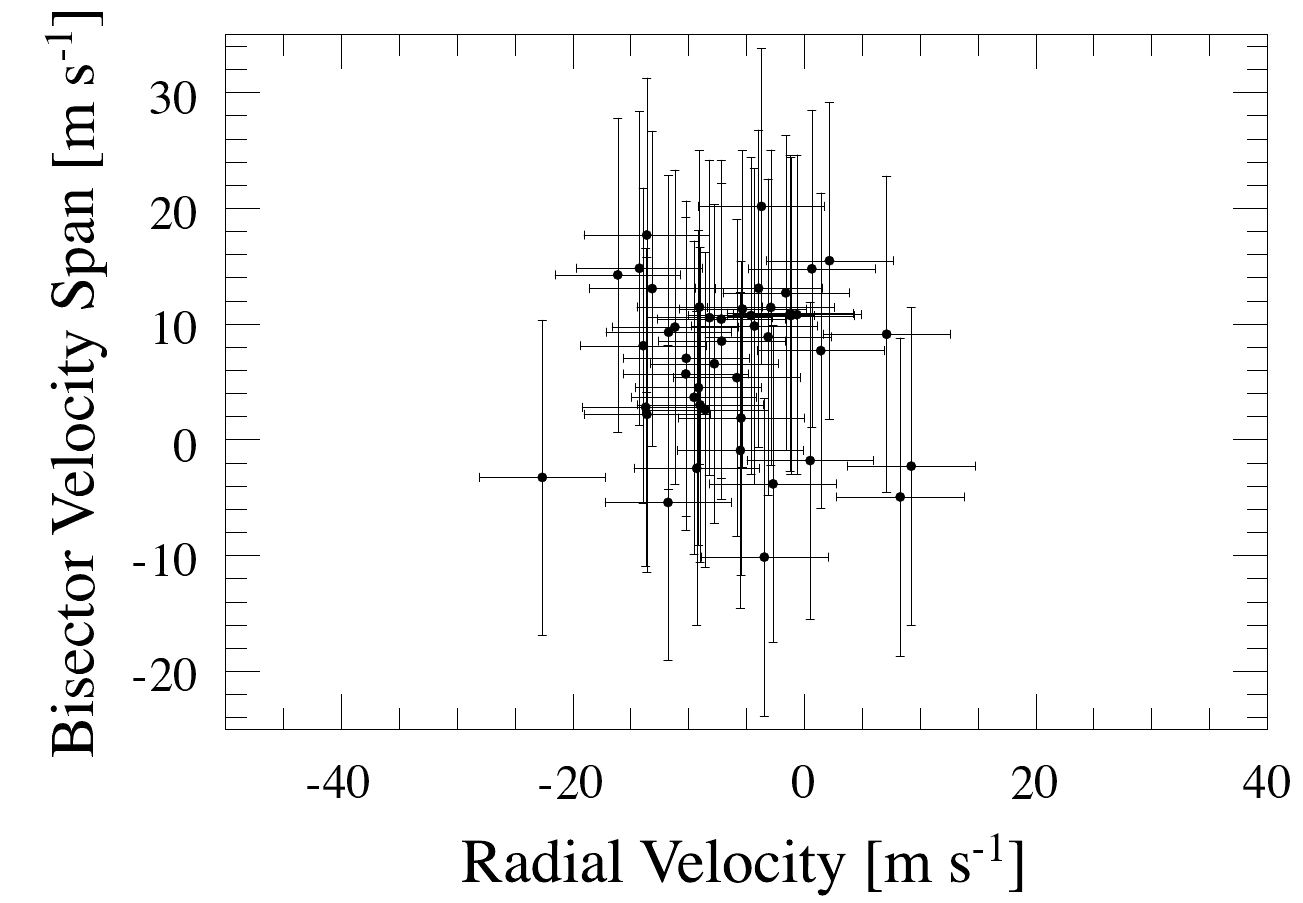}
    \caption
	{
	  {\it Top}: RVs for 1-hour \sophie~data.
	  {\it Bottom}: Bisector velocity spans versus RVs for 1-hour \sophie~data.
	}
    \label{bis_span_1h}
  \end{figure}

 Finally, to quantify its short-term variations, we monitored \thetacyg~for 1
 hour in a row (as we did for \object{HD\,60532}, Paper~V). High-frequency
 variations are due to stellar phenomena and produce a noise (jitter) that has
 to be taken into account in the analysis of longer period variations. In
 October 2008, 46 consecutive spectra were then taken with \sophie~under average
 observation conditions (airmass below 1.2, S/N $\simeq$ 160). The resulting RVs
 and associated bisector velocity spans are presented in
 Figure~\ref{bis_span_1h}. It appears that the short-term variations can account
 for an RV amplitude of $\simeq$30\,\ms~($\sigma_{\rm rv} = 6.4$\,\ms), and that
 the total bisector velocity span amplitude over the whole \sophie~data
 (Fig.\,\ref{bis_span}, bottom) set can be explained only with those short-term
 variations. In the following attempt to analyse the high amplitude RV
 variations, we will adopt increased RV uncertainties (at least $\pm$6.4\,\ms,
 fixed to that value) in order to take this stellar jitter into account.

  \subsection{Imaging data}
  \label{sec:ao}

  We observed \thetacyg~at high angular resolution and high contrast with the
  adaptive-optics (AO) instrument PUEO (\cite{rigaut98} 1998) mounted on the
  3.6-m Canada-France-Hawaii Telescope (CFHT, USA). The near-infrared camera
  used, KIR (\cite{doyon98} 1998), has a field of view of $\sim35\arcsec \times
  35\arcsec$, with a scale of $\sim0.035\arcsec$ per pixel. We performed
  non-saturated exposures as well as 30-second saturated ones, to investigate at
  a deeper level the vicinity of the star. A special care was taken to ensure
  that the non-saturated exposures could be used as references for an accurate
  measurement of the possible companions positions relative to the central star
  and as references for the measurement of the photometric contrast between the
  star and the possible companions.

\begin{figure}[t!]
    \centering
    \includegraphics[width=1\hsize]{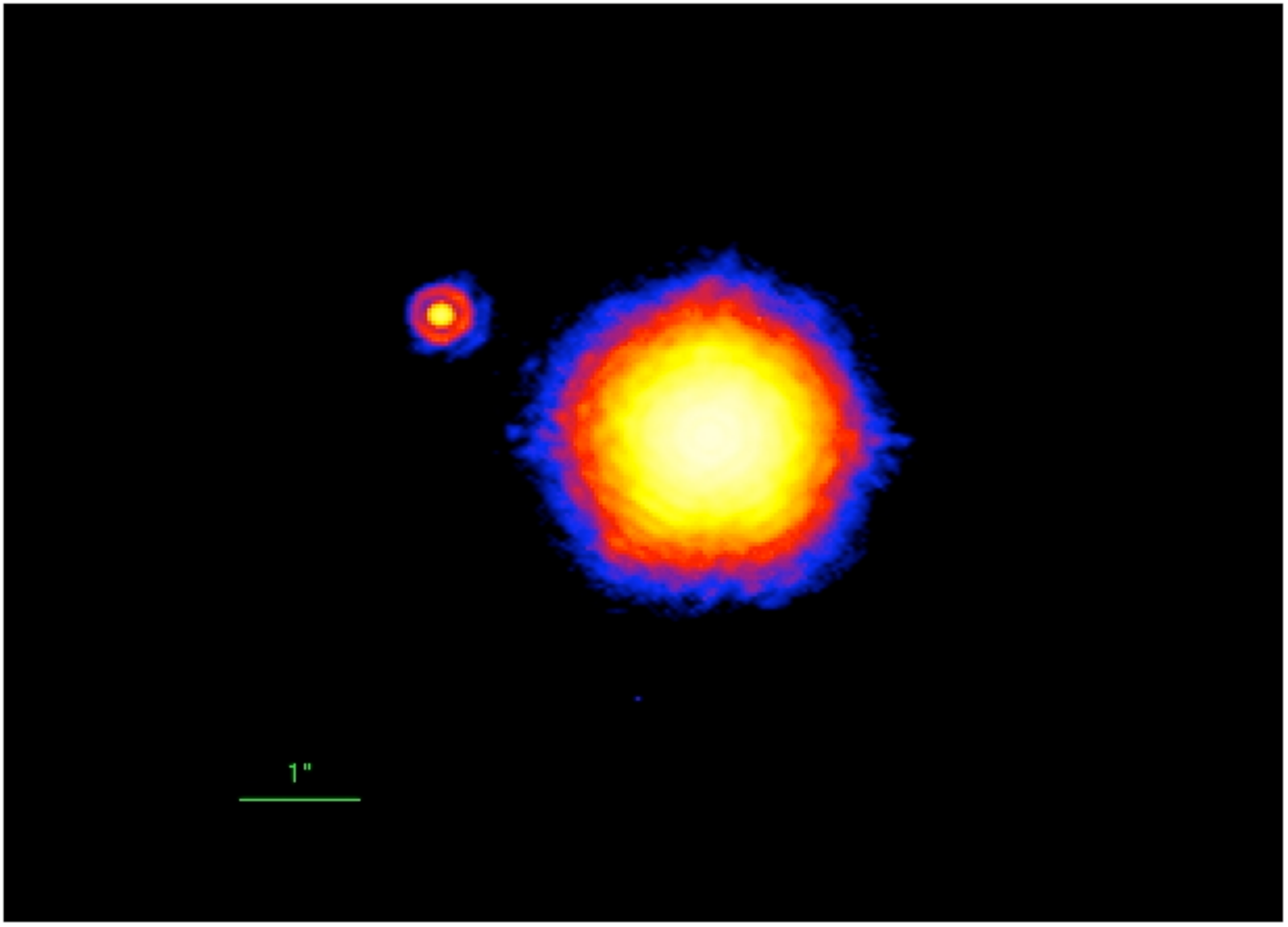}
    \caption{High contrast and angular resolution image acquired with the adaptive optics system PUEO installed at CFHT: a companion is clearly visible.
}
    \label{fig:ao}
  \end{figure}

  \thetacyg~was observed in June 2004, September 2005 and November 2007. The log
  of observations is given in Table~\ref{tab:logobs}. A classical reduction has
  been performed using the software E{\small CLIPSE} (\cite{devillard97}
  1997). A candidate companion (CC) is seen in non-saturated images (see
  Figure~\ref{fig:ao}) with the narrow bandwidth \ion{Fe}{ii} filter ($\lambda_0
  = 1.644\,\mu$m, $\Delta\lambda = 0.015\,\mu$m). A deconvolution algorithm
  using the method described in \cite{veran97} (1997) has been applied to derive
  the contrast and angular separation ($\rho$) between the star and the
  companion.

  \begin{table}[t!]
    \caption{Log of WDS observations between 1889 and 1968 followed by the AO observations using PUEO at CFHT between June 2004 and November 2007. In the case of WDS data the uncertainties are unknown.}
    \label{tab:logobs}
    \begin{center}
      \begin{tabular}{l c c c c}
	\hline
	\hline
	Date       & $\rho$           & $\theta$        & Contrast       & Band\\
                   & [\arcsec]        & [\degr]         & [mag]          &     \\
	\hline
	1889.37    & $3.62$           & $43.9$          &                &  \\
  	1892.38    & $3.79$           & $47.0$          &                &  \\
  	1898.46    & $3.37$           & $49.2$          &                &  \\
  	1898.63    & $3.71$           & $46.9$          &                &  \\
  	1958.58    & $3.42$           & $51.7$          &                &  \\
  	1968.72    & $2.92$           & $59.9$          &                &  \\
        \hline
	2004-06-28 & $2.510\pm 0.021$ & $67.37\pm 0.48$ & $4.6\pm 0.1$   & \ion{Fe}{ii}\\
  	2007-11-16 & $2.369\pm 0.005$ & $69.02\pm 0.11$ & $4.6\pm 0.1$   & \ion{Fe}{ii}\\
	\hline
      \end{tabular}
    \end{center}
  \end{table}

\begin{figure}[t!]
    \centering
    \includegraphics[width=0.9\hsize]{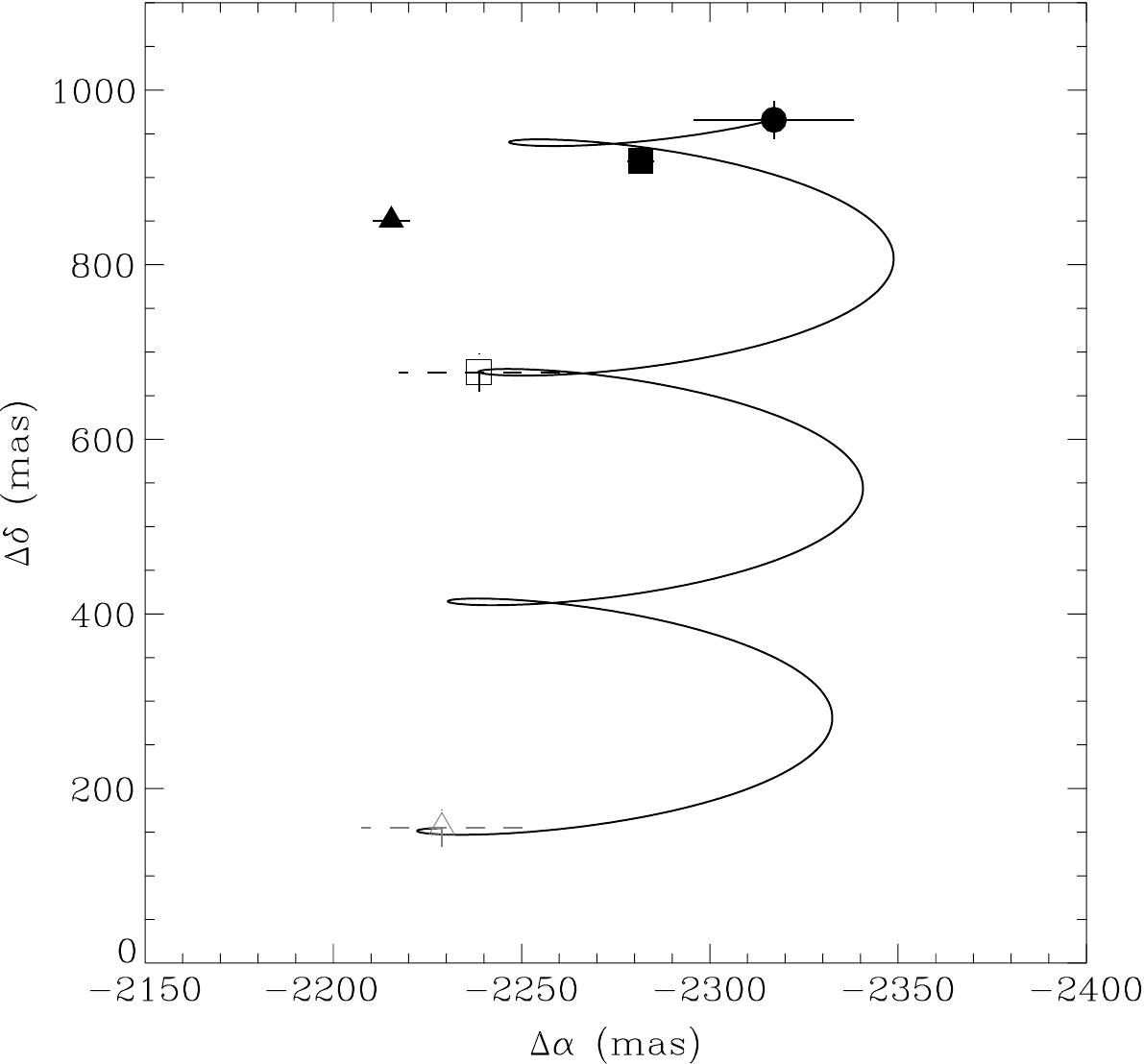}
    \caption{Temporal evolution of the separation of the stellar companion of \thetacyg~with respect to the central star. Filled symbols represent the position of the stellar companion relative to the main star for the three AO observations ({\it filled circle}: June 2004; {\it filled square}: September 2005; {\it filled triangle}: November 2007), whereas empty symbols show the positions at the same dates in case of a background star. The curve shows the path that would have followed the candidate companion if it was a background star, taking the star proper- and parallactic-motion into account.}
    \label{ppm}
  \end{figure}

  Fig.\,\ref{ppm} shows the relative positions of the CC between 2004 and
  2007. Clearly, the CC is not a background star but it is bound to
  \thetacyg. Given the H{\small IPPARCOS} distance, 18.6\,pc, we derive a
  projected separation of 46.5\,AU between the two objects, thus a minimum
  period of roughly 230 years, assuming a circular orbit. Moreover, we see in
  Fig.\,\ref{ppm} that the orbit of the companion is --- still very partially
  --- resolved over a three year period of observation.

  The measured contrast between \thetacyg~and its companion \thetacyg~B is $4.6
  \pm 0.1$\,mag in $H$ band; in $K$ band, the measured contrast is $4.5 \pm
  0.1$\,mag. Given the star apparent magnitudes provided by \cite{2mass06}
  (2006), and distance (see above), we deduce $H$ and $K$ absolute magnitudes of
  $7.0 \pm 0.1$ and $6.7 \pm 0.1$\,mag respectively. Using the BCA98
  evolutionary models (\cite{baraffe98} 1998), and assuming any age above
  $\sim$100\,Myr, we deduce a mass $m_2 \simeq$\,0.35\,\Msun~for the companion
  (the evolutionary effects are negligible). Using the empirical relation given
  by \cite{delfosse00} (2000) for $M_H \simeq 7$, we find a comparable mass $m_2
  \simeq$\,0.33\,\Msun~for the companion.

\begin{figure}[t!]
    \centering
    \includegraphics[width=0.8\hsize]{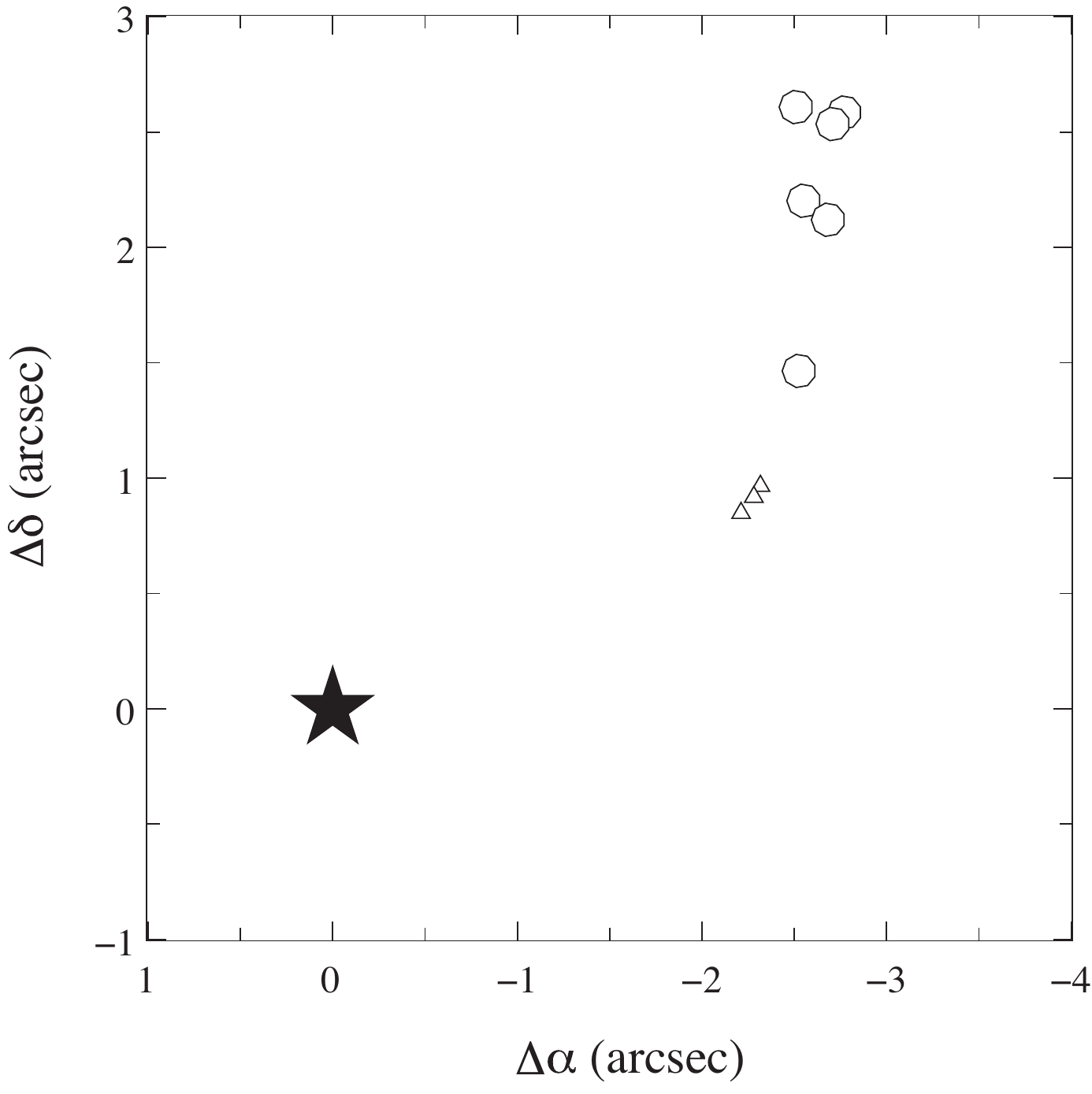}
    \caption{Positions of the companion star with respect to the primary star for more than one century. The star represents the primary \thetacyg, the circles the WDS data (uncertainties are unknown), and the triangles our AO data (uncertainties are smaller than the symbols).}
    \label{fig:imagery}
  \end{figure}

  \thetacyg~was classified as a double star (\cite{ccdm94} 1994). The Washington
  Double Star (WDS) data (\cite{hartkopf01} 2001) indicate a visual companion
  detected several times since 1889, with a magnitude of $\sim$12, \ie
  comparable to the visual magnitude expected from a 0.35-\Msun~star at
  \thetacyg~distance. The relative position of this object varies between 1892
  and 1968 (see Tab.\,\ref{tab:logobs}), and indicates that this companion is
  bound to \thetacyg. The companion was not detected by H{\small IPPARCOS}
  because the contrast with \thetacyg~was too high. It is reasonable to say that
  the WDS companion and the one found with PUEO is the same. Then we can track
  its motion over more than a century (Fig.\,\ref{fig:imagery}), but its orbit
  is still very incomplete.


\section{Origins of the observed RV variations}
\label{keplerian_sol}

We investigate hereafter different possible origins for the RV variations:
stellar phenomenon (spots, pulsations), and planets. Beforehand, we estimate the
possible impact of the stellar companion on the spectroscopic data.

  \subsection{Impact of the stellar companion on the RV data}

  \thetacyg~B could a priori play a role on the measured radial velocities in
  two ways. Firstly, given their separation ($\simeq$2\arcsec), and given the
  usual seeings and the entrance width of the optical fiber, the spectra of the
  two stars are superimposed in the \elodie~or \sophie~data (fiber entrance of
  2\arcsec~and 3\arcsec~respectively). This could introduce a bias in the
  measurements. However, the contrast of 4.6 magnitude in $H$ band translates to
  a contrast of 7.9\,mag (a flux ratio of $\sim$\,1500) in $V$ band. Hence, the
  signal of the secondary is negligible in our spectra (we expect a
  radial-velocity effect below 1\,\ms). If very actuve, it could still produce a
  weak H$\alpha$ signature superimposed on the \thetacyg~H$\alpha$ line. This
  does not affect our results as this line is not taken into account for the RV
  measurements.

  With the classical cross-correlation technique, the potential pollution of the
  spectrum by a stellar companion can be tested using various masks. We
  therefore checked that the RV amplitudes remain identical when using various
  masks.  Also, the RVs are identical if we use either the red part or the blue
  part of the spectra to measure them. This confirms that the spectrum of the
  companion has no impact on the measured radial velocities.

\begin{figure}[t!]
    \centering
    \includegraphics[width=1.\hsize]{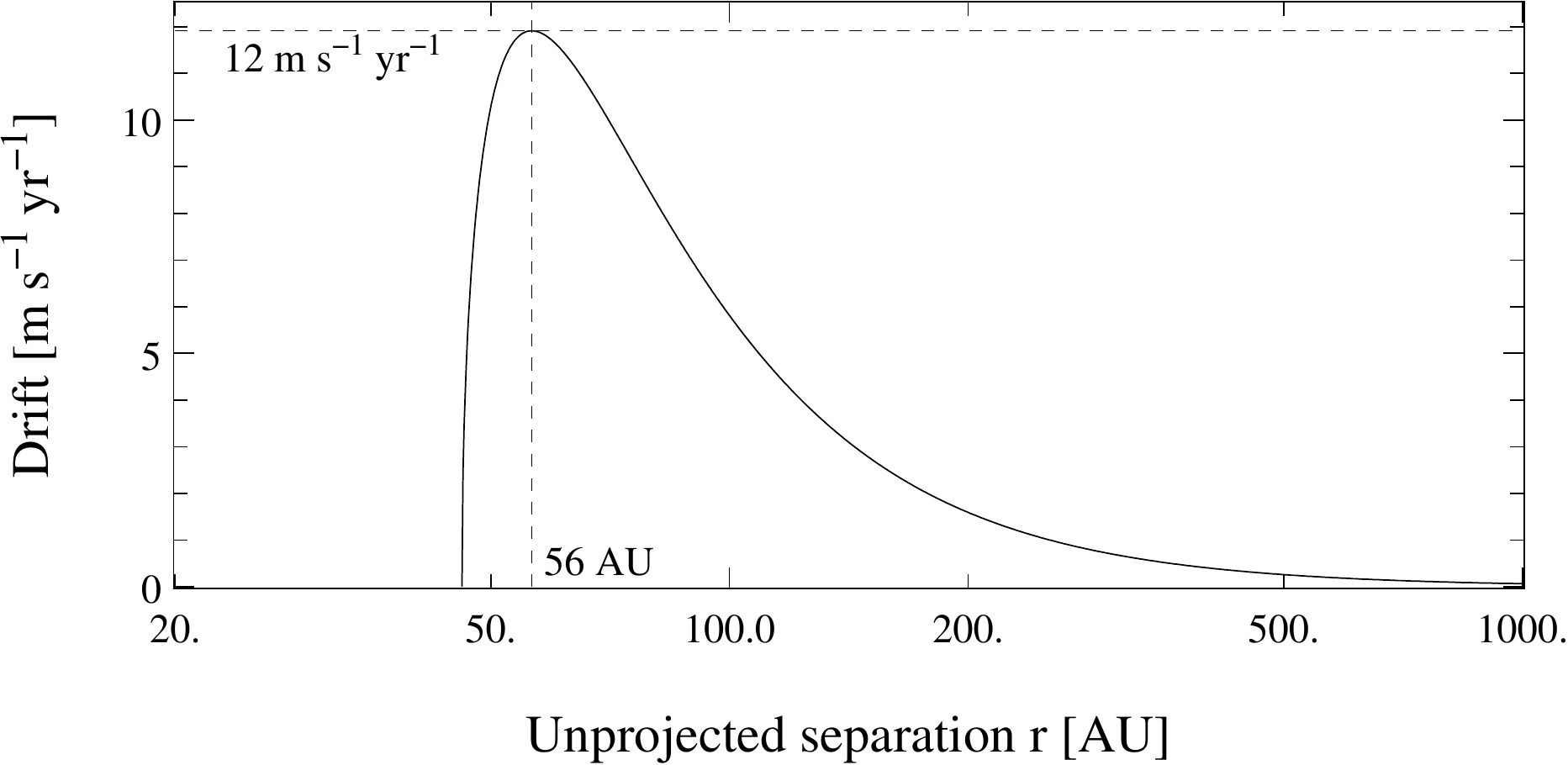}
    \caption{Possible drift induced by the stellar companion given the unprojected true separation $r$ to the primary star.}
    \label{fig:drift}
  \end{figure}

  Secondly, the stellar companion induces of course radial velocity variations
  of the primary. Given the companion properties, and assuming the system seen
  edge-on, we estimate the maximum drift possibly induced on the primary star by
  plotting the acceleration projected on the line of sight $z$, with respect to
  the true separation $r$ (Fig.\,\ref{fig:drift})
  $$\frac{d^2 z}{dt^2} = \frac{G m_2}{r^2} \cos \big[ \arcsin\big(
    \frac{\rho}{r}\big) \big] ,$$ were $G$ is the gravitational constant, $m_2 =
    0.35$\,\Msun~the mass of the secondary, and $\rho = 46$\,AU the closest
    projected separation measured. We find a maximum drift of 12\,\ms\,yr$^{\rm
    -1}$, which would lead to a maximum drift of 60\,\ms~over 5 years. In fact,
    a similar drift has to be included into our fit of the radial-velocity curve
    (see Section~\ref{oneplanet}). \thetacyg~B could then explain such a drift,
    but of course the observed periodic radial-velocity variations with an
    amplitude larger than 150\,\ms~are not explained by the presence of this
    stellar companion.

\subsection{Stellar phenomenon}

\subsubsection{Stellar spots}

  We saw in the previous section that there is no correlation between the star
  RV variations and the BVS variations. Given the star projected
  rotational-velocity, the instrument resolution, and according to the study
  presented in \cite{desort07} (2007), we can definitely conclude that the
  observed RV variations are not due to stellar spots. Indeed, would this be the
  case, a correlation between the BVS and the RV would be observed. Typically,
  given the star's properties, one or more spots on a inclined star would be
  needed to reproduce a periodic signal, and a linear correlation coefficient
  $\simeq-0.5$ between the bisector velocity spans and the RVs. Given the
  observed RV amplitude, the amplitude of the BVS would therefore be much higher
  than what is actually observed.

  Also, in such a case, one would expect significant photometric variations. A
  single spot producing such an RV variation would induce a photometric
  amplitude between 5 and 30\,mmag, depending mainly on the star inclination and
  the spot location (\cite{desort07} 2007). The photometry given by H{\small
  IPPARCOS} (\cite{hipparcos97} 1997) is constant with a scatter of only
  0.004\,mag. We recognize however that the H{\small IPPARCOS} data were not
  recorded simultaneously with the spectroscopic ones, so this photometric
  argument is certainly weaker than the absence of correlation between RV and
  BVS variations.

\begin{figure}[t!]
    \centering
    \includegraphics[width=0.8\hsize]{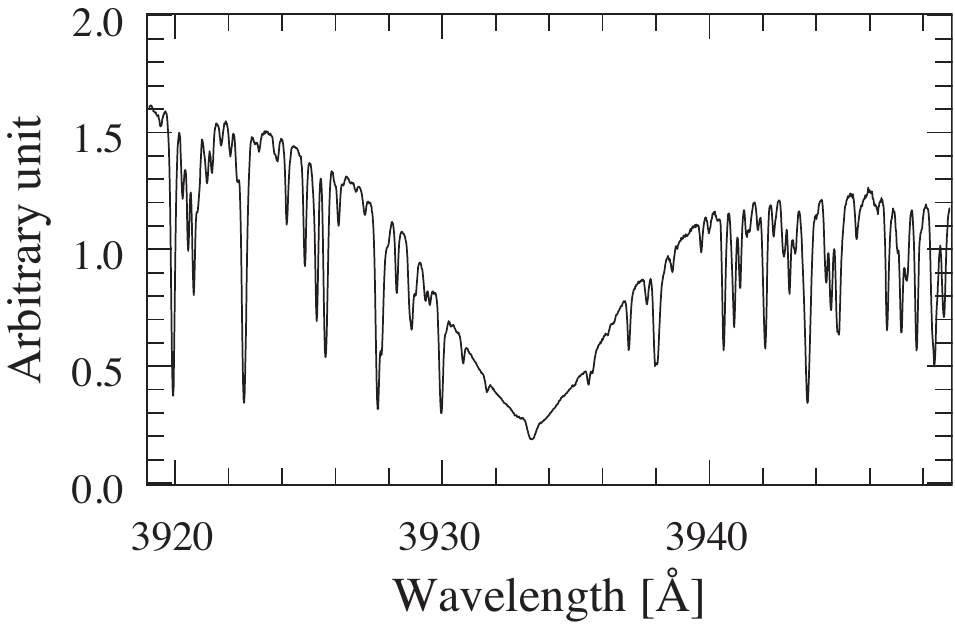}
    \caption{No emission is observed in the \ion{Ca}{ii}\,K line for the \thetacyg~spectra. This is the average spectrum of all the \sophie~spectra used, after recentering by the RV variations measured.}
    \label{CaK}
  \end{figure}

  No clear emission in the core of the \ion{Ca}{ii} lines is observed (see
  Fig.\,\ref{CaK} for the \ion{Ca}{ii}~K line); this excludes a high level of
  activity.

  We looked for possible long-term stellar variations using classical
  H$\alpha,\beta,\gamma$ indicators. Very faint H$\alpha$ variations are
  detected, but they are not correlated with the RV variations. As the H$\alpha$
  indexes are moreover quite sensitive to pollution by the thorium lamp, we
  cannot attribute them a stellar origin.

  Finally, the stellar rotational period is less than 7~days according to its
  \vsini~and assuming a stellar radius typical for this type of star. Hence a
  spot or a set of spots can not explain RV variations with periods of one
  hundred days or more.

  We can therefore safely conclude that spots are most probably not responsible
  for the observed RV variations.

\subsubsection{{\bf Granulation}}
{\bf

  Another source of variation could be the attenuation or suppression of the
  convective blueshift due to the presence of plages (\eg \cite{deming94} 1994,
  \cite{marquez96} 1996).

  The expected convective blueshift expected for \thetacyg~would be in the range
  400--800\,\ms~depending on the line (see \cite{gray09} 2009 for a F5IV--V,
  which is the closest to \thetacyg~in his sample). The observed variations
  could then be due to a cyclic variation with plages filling factor varying
  between 0 and 30\% typically in order to produce the
  220\,\ms~variation. However, such a variation of the convection properties
  should lead to a strong variation of the bissectors and of the \ion{Ca}{ii}
  index, which is not the case.

}

\subsubsection{Stellar pulsations}

  Generally, pulsations induce line-profile variations which strongly affect the
  bisector velocity span (Paper~VI). Besides, the timescale of the observed
  radial-velocity variations ($\geq$\,100 days) is far larger than the ones of
  pulsations known for this type of main-sequence stars (it is in fact more
  characteristic of giant stars variability). Variability periods of a few days
  are observed in the case of the pulsating $\gamma$~Doradus stars
  (\cite{mathias04} 2004). Moreover, if we integrate the RVs measured between a
  minimum and a maximum of the amplitude for half a period, we end up with a
  total displacement close to the stellar radius, which, if even possible, would
  lead to detectable photometric variations.

  It is therefore unlikely that classical pulsations are responsible for the
  observed RV variations. However, the presence of a peak at about 150 days in
  the BVS periodogram and in the RV periodogram indicates that the period of the
  RV variations is linked in some way to the low amplitude line shape
  variations. Such a situation has never been reported to our knowledge and is
  indeed quite puzzling. We cannot at this stage exclude that we could be facing
  a new type of stellar variability, undetected so far because of a lack of
  long-term, very-precise RV monitoring of main-sequence stars.

  \subsection{Planet(s) around \thetacyg}
  \label{oneplanet}

  \thetacyg~RV curve shows quasi-periodic variations with a period of
  $\sim$130--150 days, together with a variable positive trend. Moreover, the RV
  curve seems to be modulated in amplitude. We now try to see whether these
  variations can be attributed to a planet or a planetary system.

\begin{figure}[t!]
      \centering
      \includegraphics[width=1\hsize]{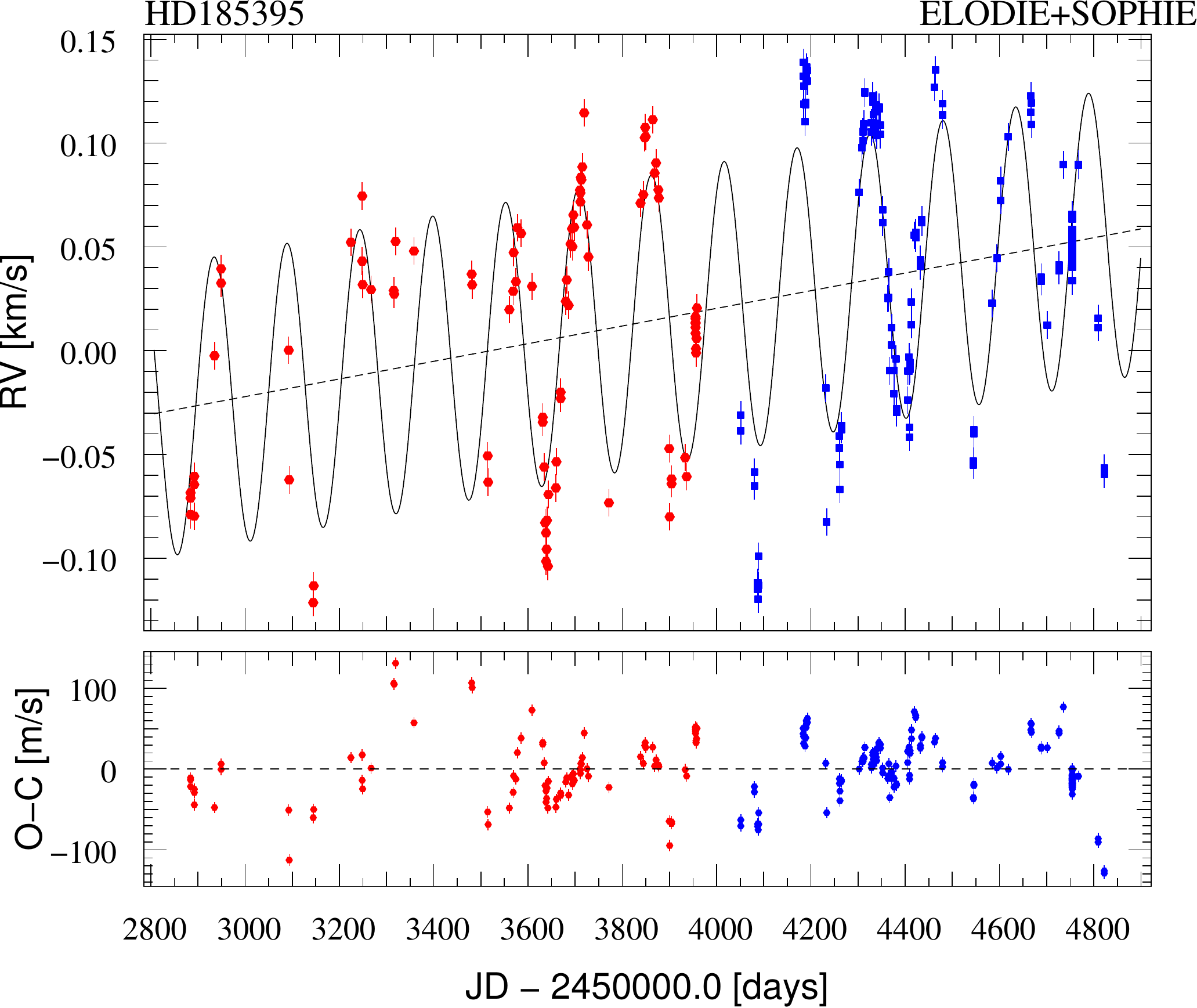}
      \caption{\elodie~and \sophie~radial velocities and orbital solution for \thetacyg, with one planet (2.3\,\Mjup~at $\sim$0.6\,AU, circular orbit) and a drift. The residuals to the fitted orbital solution are displayed below.}
      \label{fit_1CC_drift}
    \end{figure}

  We first try to fit the whole data set (with uncertainties set to
  $\pm$6.4\,\ms~) assuming a single planet and allowing a drift. We fail to
  reproduce correctly the observed RVs.  Figure~\ref{fit_1CC_drift} shows an
  example of fit with a planet on a circular orbit (plus a drift). The planet
  that produces such an RV curve has a mass of 2.3\,\Mjup, and is on a 0.6\,AU
  circular orbit (Table~\ref{tab_1CC_drift_param}). The algorithm ends up with a
  period of 155 days hung on the \sophie~data set, but fails to fit properly the
  \elodie~data set, as if there was period/phase change with time (clearly
  visible as structure in the residuals near the range [3600--3800] days). Note
  that the additional drift needed is 16\,\ms\,yr$^{\rm -1}$, slightly higher
  than the maximum value that the binary companion would probably produce. We
  note that such a system which fails to fit satisfactorilly the {\it whole }
  set of data (taken over more than 5 years) would allow to fit the data if they
  were limited to one or two consecutive periods.

\begin{figure}[t!]
      \centering
      \includegraphics[width=1\hsize]{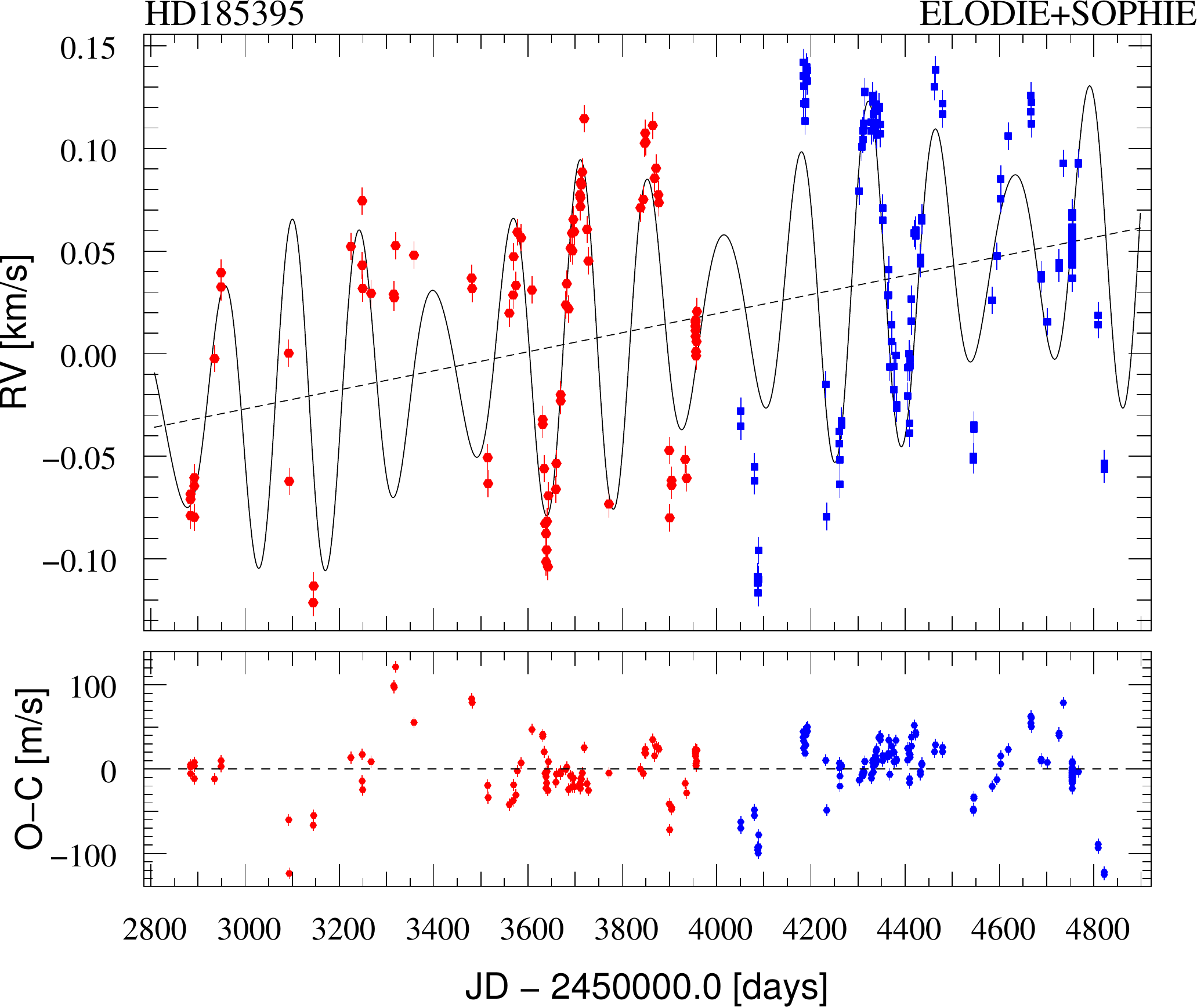}
      \caption{\elodie~and \sophie~radial velocities and orbital solution for \thetacyg, with two planets (2.1\,\Mjup~at $\sim$0.6\,AU and 0.8\,\Mjup~at $\sim$0.5\,AU on circular orbits) and a drift. The residuals to the fitted orbital solution are displayed below.}
      \label{fit_2CC_drift}
    \end{figure}

  We then tried to fit simultaneously the whole set of data, assuming the
  presence of several planets, and using a genetic algorithm search. No stable
  solution was found. The only satisfactory fits are achieved by unstable
  systems with orbits that cross each other. We show an example in
  Fig.\,\ref{fit_2CC_drift} of a fit obtained with a two-companion plus drift
  model. The residuals are still very high and the fit is not improved compared
  to the one-companion plus drift model (residual rms 35\,\ms~versus 39\,\ms,
  and the drift that we get is approximately the same: 17\,\ms\,yr$^{\rm
  -1}$~versus 16\,\ms\,yr$^{\rm -1}$). Moreover, such kinds of configurations
  with massive planets on so close orbits are not dynamically stable.

  The system is then obviously more complex than just consisting of one, two or
  even three planets plus a drift. We explored then more exotic configurations:

  \begin{itemize}

  \item Instead of harbouring one single planet, the system could consist of a
    binary planet system orbiting the star, much like the Pluto--Charo system,
    but with larger masses. However, the radial-velocity signal generated by
    this configuration would be very close to the one generated by a single
    planet having the total mass. The only departure from the pure Keplerian
    system would be related to the secular perturbations of the orbit due to the
    binary nature of the system. The associated secular period would be at least
    several hundreds of primary orbital periods, \ie much longer than our
    observation time span. Hence we should mainly detect the primary orbital
    signal with good accuracy.

  \item One could also think of 2 co-orbiting planets locked in 1:1 mean-motion
    resonance. Such a configuration has been observed in the system of Saturn
    satellites. The two satellites orbit Saturn on the same orbit. In the
    rotating frame, their have synchronised horseshoe-like libration motions
    that prevent them to collide.  The less massive one has the largest
    amplitude motion (\cite{yoder83} 1983).  The associated libration period is
    here again a few hundreds of orbital periods of the primaries. So, over a
    smaller time span, the radial-velocity signal of the whole system will
    mainly consist of the sum of the individual signals of the two planets. If
    we assume that the two planets have zero eccentricities, the radial velocity
    signal will have the form $A\cos(nt)+B\cos(nt+\phi)$, where $n$ is the
    common mean motion of the two planets, $A$ and $B$ are amplitudes related to
    the masses of the planets, and $\phi$ is a phase shift that depends on their
    current mutual configuration. This can be rewritten as $C\cos(nt+\psi)$
    where $C$ and $\psi$ are new amplitudes and phases that depend on $A$,$B$
    and $\phi$. This is equivalent to the signal generated by one single
    planet. This remains true if the planets have small eccentricities; and if
    they have larger eccentricities, the mutual system is not stable. Here
    again, the only departure from this signal will be the temporal variation of
    $C$ and $\psi$ that is related to the mutual libration motion of the two
    bodies. Hence we should not expect to see changes before several
    decades. The signal detected over our observation timespan should not be
    different to that of a single planet.

  \end{itemize}

    \begin{table}[t!]
      \caption{\elodie/\sophie~best orbital solution for \thetacyg, considering one planet and a drift.}
      \label{tab_1CC_drift_param}
      \begin{center}
	\begin{tabular}{l l c}
	  \hline
	  \hline
	  Parameter           &                      & \thetacyg~{\it b} \\
	  \hline
	  $P$                 & [days]               & $154.5 \pm 0.4$  \\
	  $T_0$               & [JD$-$2450000]       & $4016 \pm 1$ \\
	  $e$                 &                      & 0 (fixed)    \\
	  $\omega$            & [deg]                & 0            \\
	  $K$                 & [\ms]                & $70 \pm 4$   \\
	  $N_{\rm meas}$      &                      & 253          \\
	  $\sigma_{O-C}$      & [\ms]                & 38.7         \\
          reduced $\chi^2$    &                      & 6.1          \\
	  \hline
	  $a_1\sin{i}$        & [$10^{-3}$\,AU]      & 0.99         \\
	  $f(m)$              & [$10^{-9}$\,\Msun]   & 5.5         \\
	  $M_1$               & [\Msun]              & 1.38         \\
	  $m_2\sin{i}$        & [\Mjup]              & 2.29         \\
	  $a$                 & [AU]                 & 0.63         \\
	  \hline
	  drift               & [\ms\,yr$^{\rm -1}$] & $16 \pm 4$   \\
	  \hline
	\end{tabular}
      \end{center}
    \end{table}

\section{Concluding remarks}

  The radial velocities obtained with \elodie~and \sophie~on \thetacyg~are
  quasi-periodically variable, with a $\simeq$ 150 days period. We have
  investigated different possible origins to these RV variations, either star or
  planet related.

  Given today knowledge on stellar activity, using several criteria as usually
  done in this type of studies, we fail to attribute these variations to the
  star itself (spots, pulsations). However the presence of a peak in the
  periodogram of the BVSs at $\simeq$ 140 days, \ie close to the period of the
  RV variations prevents us from excluding totally a stellar origin.

  We tried to fit the data with one planet orbiting at less than 1\,AU and with
  a mass of a few Jupiter masses, taking also into account the impact of the
  observed companion star. It appears that such an hypothesis allows us to fit
  only part of the data, recorded over a limited number of periods, but fails to
  fit satisfactorilly the whole set of data (taken over more than 5 years). More
  complex systems were investigated, but no convincing result was obtained. More
  observations and detailed studies of gravitational interaction between the 2
  planets are needed to understand this sytem.

  With the data available, we are then not able to conclude on the origin of
  these puzzling RV variations.

  If a planet origin is confirmed, then \thetacyg, with a spectral type of F4V,
  would be the earliest main-sequence star hosting planets found so
  far. Moreover, it would be one of the few low metallicity stars hosting
  planets. Its planetary system would not be simple, and would in particular
  include strongly interacting planets.

  If a stellar origin was to be confirmed, then this would show that,
  unexpectedly, some main-sequence stars, not classified as active from usual
  criteria (\eg \ion{Ca}{ii}\,H\&K indexes) or pulsating, may undergo intrinsic
  variations which produce quasi periodical, large-amplitude and long-period
  (more than 100 days) RV variations, with at the same time, low levels of line
  shape deformations (hence small amplitude bisector velocity span
  variations). Such situations have not been considered so far in the analysis
  of RV variations, and would need to be considered in future searches for such
  long period planets, both in terms of observational strategy and data
  analysis.

  In any case, we can conclude that \thetacyg~is an individual complex system
  that deserves much more observations to be understood, and that it may also
  serve as an example for other searches.

\begin{acknowledgements}

  We acknowledge support from the French CNRS and the support from the Agence
  Nationale de la Recherche (ANR grant NT05-4\_44463). We are grateful to the
  Observatoire de Haute-Provence (OHP) and the CFHT for their help during the
  observations, and to the Programme National de Plan\'etologie (PNP, INSU).
  
  These results have made use of the SIMBAD database, operated at CDS,
  Strasbourg, France. They also make use of data products from the Two Micron
  All Sky Survey, which is a joint project of the University of Massachusetts
  and the Infrared Processing and Analysis Center/California Institute of
  Technology, funded by the National Aeronautics and Space Administration and
  the National Science Foundation.

  We also thank G\'erard Zins and Sylvain C\`etre for their help in implementing
  the SAFIR interface.

  X.B. acknowledges support from the Funda\c{c}\~ao para a Ci\^encia e a
  Tecnologia (Portugal) in the form of a fellowship (reference
  SFRH/BPD/21710/2005) and a programme (reference PTDC/CTE-AST/72685/2006), as
  well as the Gulbenkian Foundation for funding through the ``Programa de
  Est\'imulo \`a Investiga\c{c}\~ao''.

  N.C.S. would like to thank the support from Funda\c{c}\~ao para a Ci\^encia e
  a Tecnologia (Portugal) in the form of a grant (references
  POCI/CTE-AST/56453/2004 and PPCDT/CTE-AST/56453/2004), and through programme
  Ci\^encia\,2007 (C2007-CAUP-FCT/136/2006).

\end{acknowledgements}

\end{document}